\documentclass[prb, twocolumn, reprint, superscriptaddress, aps, longbibliography,
floatfix, reprint]{revtex4-2}

\usepackage[utf8]{inputenc}
\usepackage[T1]{fontenc}

\usepackage{graphicx}
\usepackage{amsmath}
\usepackage{amssymb}
\usepackage{afterpage}
\usepackage{amsfonts}
\usepackage{braket}
\usepackage{textcomp}
\usepackage{float}
\usepackage{bm}
\usepackage{siunitx}
\usepackage{lipsum}
\usepackage[]{xcolor}
\usepackage{hyperref}
\hypersetup{
	colorlinks,
	linkcolor={blue!90!black},
	citecolor={blue!90!black},
	urlcolor=	{blue!90!black}
}
\usepackage[]{placeins}

\usepackage[]{newtxtext} 
\usepackage[subscriptcorrection,nosymbolsc,smallerops,bigdelims]{newtxmath} 
\DeclareMathAlphabet{\mathcal}{OMS}{cmsy}{m}{n} 
\DeclareMathAlphabet{\mathbcal}{OMS}{cmsy}{b}{n} 

\begin{document}
\title{Quantum dots as optimized chiral emitters for photonic integrated circuits}

\author{Jakub Rosi\'n{}ski}
\author{Micha\l{} Gawe\l{}czyk}
\affiliation{Institute of Theoretical Physics, Wroc\l{}aw University of Science and Technology,  50-370 Wroc\l{}aw, Poland}
\author{Karol Tarnowski}
\affiliation{Department of Optics and Photonics, Wroc\l{}aw University of Science and Technology, 50-370 Wroc\l{}aw, Poland}
\author{Pawe\l{} Karwat}
\author{Daniel Wigger}
\affiliation{School of Physics, Trinity College Dublin, Dublin 2, Ireland}
\author{Pawe\l{} Machnikowski}
\affiliation{Institute of Theoretical Physics, Wroc\l{}aw University of Science and Technology, 50-370 Wroc\l{}aw, Poland}



\begin{abstract}
Chiral coupling, which allows directional interactions between quantum dots (QDs) and photonic crystal waveguide modes, holds promise for enhancing the functionality of quantum photonic integrated circuits. Elliptical polarizations of QD transitions offer a considerable enhancement in directionality. However, in epitaxial QD fabrication, the lack of precise control over lateral QD positions still poses a challenge in achieving efficient chiral interfaces. Here, we present a theoretical analysis in which we propose to optimize the polarization of a QD emitter against the spatially averaged directionality and demonstrate that the resulting emitter offers a considerable technological advantage in terms of the size and location of high-directionality areas of the waveguide as well as their overlap with the regions of large Purcell enhancement, thereby improving the scalability of the device. Moreover, using ${\bm k}\cdot{\bm p}$ modeling, we demonstrate that the optimal elliptical polarization can be achieved for neutral exciton transitions in a realistic QD structure. Our results present a viable path for efficient chiral coupling in QD-based photonic integrated circuits, to a large extent overcoming the challenges and limitations of the present manufacturing technology.
\end{abstract}


\date{\today}

\maketitle

\section{Introduction}
Applications in quantum information processing and quantum communication have attracted much attention to quantum networks based on on-chip photonic integrated circuits~\cite{o2009photonic, lodahl2017quantum, wang2020integrated}. A fundamental requirement for storing, processing, and exchanging information is to interface quantum nodes with quantum channels on such physical platforms~\cite{cirac1997quantum, yao2005theory, tanzilli2005photonic, ritter2012elementary, kalb2017entanglement,lodahl2015interfacing, chan2022quantum}. This requires emitters that couple to photonic pathways, encoding the quantum information of a long-lived qubit onto a flying transporting qubit. On the one hand, in the past few years, several nanophotonic waveguide schemes have been proposed as channels, including photonic crystal waveguides (PCWs)~\cite{sollner2015deterministic,mahmoodian2016quantum, hauff2022chiral, siampour2023observation, martin2023topological}, optical fibers~\cite{petersen2014chiral, mitsch2014quantum}, crossed nanowire waveguides~\cite{luxmoore2013interfacing, luxmoore2013optical}, and nanobeam waveguides~\cite{coles2016chirality,coles2017path,mrowinski2019directional}, which utilize quantum interfaces to transfer data from matter to photonic qubits. On the other hand, self-assembled semiconductor quantum dots (QDs) constitute an interesting class of emitters for photonic quantum technology applications~\cite{wang2019boson, loredo2017boson}. QDs may be easily integrated into photonic nanostructures, since they are solid-state emitters and the spins in QDs have been identified as long-lived qubits~\cite{imamog1999quantum, warburton2013single, yoneda2018quantum, norman2018perspective}.

In a QD-based implementation, quantum information has to be transferred to photonic channels by a controlled interface between the photonic propagating mode and the charge or spin states of the QD. In a perfect world, such an interface would be deterministic, meaning that emission fully takes place into the desired mode, and the information is transmitted without information backflow or loss. Chiral interfaces that support directional interactions represent a promising paradigm for deterministically transferring the quantum state from the solid-state platform to the quantum state of light, exploiting the direction-dependent nature of light-matter interaction~\cite{lodahl2017chiral}. The directional mode excitation arises from the longitudinal component of the electric field~\cite{lax1975maxwell, bliokh2012transverse} which, due to the $\pm \pi/2$ phase shift with respect to the transverse field component, makes the polarization elliptical in the propagation plane. This polarization flips its rotation direction with the inversion of the propagation direction due to the time-reversal symmetry of Maxwell’s equations~\cite{bliokh2015transverse, bliokh2015spin, aiello2015transverse}. Chiral coupling emerges naturally in nanophotonic systems, including plasmonic structures~\cite{lee2012role, rodriguez2013near, lin2013polarization}, crossed nanowire waveguides~\cite{luxmoore2013interfacing,luxmoore2013optical}, whispering-gallery-mode resonators~\cite{junge2013strong}, dielectric nanobeam waveguides~\cite{coles2016chirality, siampour2023observation}, nanofibers~\cite{mitsch2014quantum, sayrin2015nanophotonic}, PCWs~\cite{sollner2015deterministic, le2015nanophotonic, coles2016chirality, scarpelli201999, hauff2022chiral, siampour2023observation, martin2023topological}, and topological PCWs~\cite{barik2018topological, mehrabad2020chiral, hauff2022chiral, siampour2023observation, martin2023topological}. 

The coupling depends both on the polarization of the mode propagating in the photonic structure at the position of the emitter and on the polarization of the transition dipole moment of the emitter itself, which opens a twofold way for optimization. On the one hand, in recent years PCWs have been demonstrated as promising structures offering the possibility of engineering local polarization and dispersion~\cite{burresi2009observation, sollner2015deterministic, young2015polarization, mahmoodian2017engineering, Lang_2017, beggs2017optimised}. On the other hand, it has been observed that, by properly engineering the polarization of the emitter, one attains an additional degree of control over the directionality of the emission
~\cite{rodriguez2013near}. In particular, elliptically polarized emitters coupled to photonic waveguides show an advantage over circular ones with respect to the directionality and efficiency of the coupling ~\cite{lang2022perfect}. This results from the fact that the guided mode is circularly polarized only in close proximity of a few points in the PCW unit cell (C points)~\cite{lang2015stability}, while elliptical polarization states predominate throughout the majority of the unit cell~\cite{burresi2009observation} thus restricting the area where the chirality of the coupling to a circular dipole is optimal and limiting the overlap between the areas of high directionality and large Purcell enhancement. 

It has been suggested ~\cite{lang2022perfect} that QDs can be the optimal emitters, taking advantage of their elliptical polarization resulting from hole band mixing \cite{Koudinov2004,MusialPRB2012}. Indeed, in epitaxial nanostructures, the reduced symmetry of the confinement potential, which is influenced by in-plane geometry, strain anisotropy, or atomistic asymmetry, leads to mixing between heavy and light holes in the valence band. As a result, bright electron-hole configurations of QDs are associated with elliptical dipoles with typical degrees of linear polarization of 1\%–20\% ~\cite{MusialPRB2012}. For excitonic transitions (creation or recombination of an electron-hole pair), the two elliptically polarized states are coupled by electron-hole exchange with a relative phase locked by time reversal, leading to two linearly polarized transitions. Elliptical polarization can be restored by applying a magnetic field, offering a degree of control by exploiting the competition between the electron-hole exchange and the Zeeman energies. For a trion transition (electron-hole recombination in the presence of an additional, resident carrier) the electron-hole exchange is suppressed due to the singlet configuration of the two identical carriers (electrons or holes) and the system shows two time-reversal-symmetric elliptical transitions, corresponding to the two spin orientations of the final single-particle state.  

The main drawback of QDs fabricated by conventional growth techniques is the lack of precise control over their lateral positions. QDs are fabricated using epitaxial techniques, in which heterostructures are grown in individual layers~\cite{shchukin1999spontaneous, stangl2004structural, biasiol2011compositional}. The most common approach is the Stranski-Krastanov method, which relies on the self-assembly of a QD layer on a substrate surface due to the lattice mismatch between the layers. One route to achieving control over the location of QDs is based on the site-controlled growth of QDs~\cite{schneider2008lithographic, schneider2009single, skiba2011narrow, jons2013triggered}. However, the properties of such QDs in terms of optical quality (emission linewidths and quantum efficiency) cannot match those of QDs based on Stranski-Krastanov growth~\cite{albert2010quantum, huggenberger2011narrow} yet. As a viable alternative, nanophotonic structures can be aligned with a single QD that is first located by microscopy techniques~\cite{badolato2005deterministic, hennessy2007quantum, dousse2008controlled, thon2009strong, kojima2013accurate}. However, such an integrated system cannot be scaled to more than a few QDs. Furthermore, nanofabrication protocols based on pre-located QDs allow for the emitter-photonic structure integration with an accuracy of around $40$~nm~\cite{pregnolato2020deterministic}, limited by imperfections within the imaging system and the subsequent nanofabrication. This accuracy results in a precision spot with a radius larger than the area of highly directional emission by a circular dipole~\cite{coles2016chirality}. 

Here, we follow the approach based on elliptical polarization engineering of the transition dipole moment ~\cite{lang2022perfect} and assess the suitability of QDs as engineered quantum emitters. Taking into account the random QD positions, we propose to use the position-averaged directionality of emission into a glide-plane photonic waveguide as a figure of merit. We determine this quantity as a function of the polarization of the emitter dipole, determining the optimal polarization of the emitter, which is found to yield nearly 60\% average directionality. The selected figure of merit is validated by showing that the resulting optimal polarization yields a significant technological advantage from the point of view of the required precision of placement of the emitter. Next, by quantitative modeling of QD properties using a combination of ${\bm k}\cdot{\bm p}$ and configuration-interaction methods, we show that the optimal polarization properties correspond to realistic compositional and morphological characteristics of charge-neutral QDs that emit in the technologically advantageous telecom spectral range~\cite{SauerwaldAPL2005,MusialPRB2012,GawelczykPRB2017}. 

\section{Model}\label{sec:model}
We study a coupled PCW-QD system that realizes both a strong photon-emitter interaction and high directionality. In this section, we describe the waveguide model followed by the QD model and coupling characterization. 

\subsection{Waveguide}\label{subsec:waveguide}

 \begin{figure}[tb!]
	\includegraphics[width=\columnwidth]{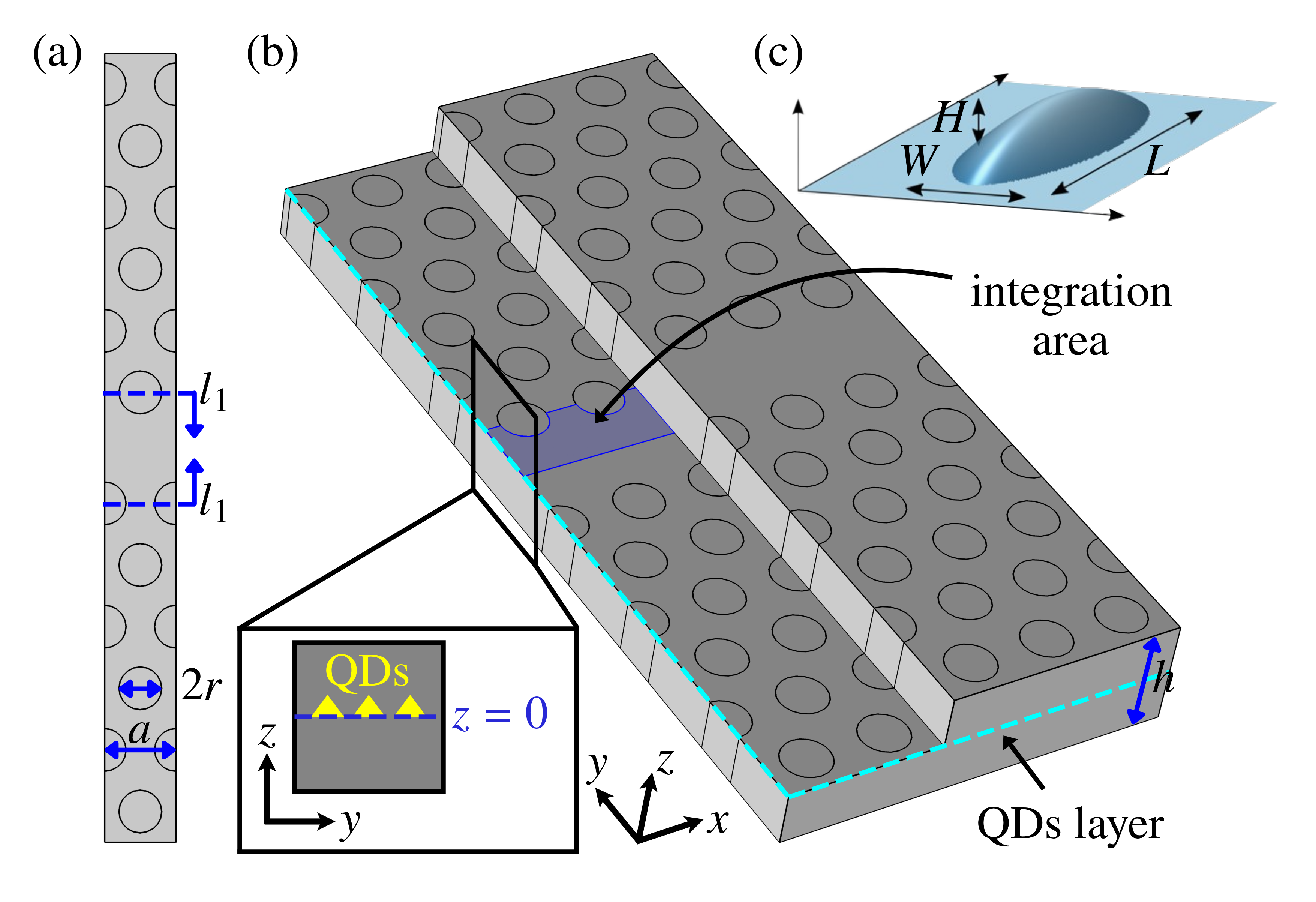}
	\caption{(a) A supercell of the PCW with the hexagonal lattice of air holes and an up-down glide-plane symmetry. The waveguide is described by the lattice constant $a$, the hole radius $r$, and the membrane thickness $h$. Additionally, the first row of holes is shifted by a distance $l_1$. (b) Perspective view of the photonic structure showing the QD layer position ($z=0$, indicated by the dashed blue line) and averaging areas (highlighted in blue shading). (c) Schematic image of the QD geometry illustrating length $L$, width $W$, and height $H$.}\label{Fig1}
\end{figure}

The geometry of the considered photonic crystal waveguide is presented in Fig.~\ref{Fig1}. The waveguide is created by a missing row of holes in a hexagonal pattern of air holes with equal radii $r=0.3a$, a membrane thickness $h=0.64a$, and a refractive index $n=3.35$, where $a$ denotes the lattice constant of the photonic structure. Furthermore, one side of the waveguide is shifted by half a lattice constant along the propagation direction, leading to a glide-plane waveguide structure. Such a symmetry breaking results in elliptically polarized light at the field maxima, effectively combining efficient chiral coupling with strong Purcell enhancement~\cite{sollner2015deterministic}. Consequently, the current research identifies glide-plane PCW as the most promising platform for chiral interactions ~\cite{hauff2022chiral, siampour2023observation, martin2023topological}. The structure is further modified by shifting the first rows of holes toward the center of the waveguide by $l_1 = a\sqrt{3}/20$. Although modifying the diameter of the air holes might bring additional advantages~\cite{schulz2010dispersion}, it is less accurately realized in electron-beam lithography than changing their position. Therefore, we restrict the current study to structures with a fixed hole size. We design the waveguide to guide light with wavelengths near $\lambda=1550$~nm (0.8~eV), i.e., tuned for the third telecommunication window with the lowest optical fiber losses. This is achieved by choosing the lattice constant to be $a= 433$~ nm. However, the waveguide design studied can be easily scaled~\cite{joannopoulos1997photonic} for use with any other quantum photonic platform. 

We perform numerical finite element calculations using the commercial software COMSOL Multiphysics ~\cite{linkComsol} to find the Bloch modes of the PCW and their eigenfrequencies. We apply the Floquet boundary conditions in the propagation ($x$) direction and periodic boundary conditions in the $y$ direction (Fig.~\ref{Fig1}b). We use an air layer of thickness $3h$ and perfectly matched layers of thickness $1.5h$ above and below the slab (in the $z$ direction), where $h$ is the thickness of the slab. We also introduce a perfect magnetic conductor plane, positioned in the reflection symmetry plane of the waveguide to limit the study to transverse electric (TE) modes and thus reduce the computational domain. We use a supercell of length $a$ along the waveguide, width $8.5a\sqrt{3}-2 l_1$, and height $5h$, which is large enough to suppress interaction with neighboring supercells. The calculations utilize a mesh with a maximum distance between the nearest mesh nodes of $0.4a/n$, where $n$ is the refractive index of the domain. This mesh ensures that the eigenvalues converge with an accuracy of less than $0.1\%$. 

Based on the amplitudes and phases for the components of the electric field, found from the simulations, we parameterize the in-plane part of the electric field as
\begin{equation}\label{eq:field}
\boldsymbol{e} = e_0
{\begin{pmatrix}
\cos(\chi_{\rm f}) \cos(\Psi_{\rm f}) - i\sin(\chi_{\rm f}) \sin(\Psi_{\rm f}) \\
\cos(\chi_{\rm f}) \sin(\Psi_{\rm f}) + i\sin(\chi_{\rm f}) \cos(\Psi_{\rm f})
\end{pmatrix}},
\end{equation}
where $e_0$ is the field amplitude, the angle $-\pi/2 \geq \Psi_{\rm f} \geq \pi/2$ describes the orientation of the polarization ellipse, and the angle $-\pi/4 \geq \chi_{\rm f} \geq \pi/4$ defines the degree of ellipticity, with $\tan\chi_{\rm f} = u/v$, where $v$ and $u$ are the major and minor semiaxes, respectively, so that $\chi_{\rm f} = 0$ corresponds to linear polarizations, $\chi_{\rm f}=\pm \pi/4$ yields two opposite circular polarizations, and intermediate values of $\chi_{\rm f}$ correspond to elliptical polarization. More details concerning the parametrization can be found in Appendix~\ref{AppElipsy}. 

\subsection{Quantum dots}\label{subsec:qd}

We consider widely studied InAs/AlGaInAs self-assembled QDs~\cite{SauerwaldAPL2005,MusialPRB2012,GawelczykPRB2017} that can emit in the third telecom window centered at 1550~nm wavelength. Such QDs are characterized by a larger than typical in-plane size and asymmetry. The latter is defined by the lateral aspect ratio, which may vary from below 2 to about 10~\cite{RudnoRudzinskiAPL2006}. The suitability of these QDs for the considered chiral interface results from the asymmetry-enabled mixing of heavy- and light-hole subbands, which leads to elliptically polarized transition dipoles~\cite{MusialPRB2012} that can be controlled not only by the morphology of the structure but also by a magnetic field applied along the QD growth axis (Faraday geometry; perpendicular to the waveguide in our case).

For our modeling, we take a typical InAs/AlGaInAs QD~\cite{SauerwaldAPL2005} placed on a $\sim 1$~nm thick InAs wetting layer. The QD has a triangular cross-section with a width-to-height ratio of $W/H = 6$. We vary $H$ from $1.2$ to 4.5~nm (as measured from the top of the $1.2$-nm-thick wetting layer) and the length $L$ from 20 to 60~nm in the simulation series, keeping the $W/H$ ratio fixed. We assume initially homogeneous composition in the QD and wetting layer (80\% InAs, partly mixed with barrier material) and then simulate the interdiffusion of material at interfaces by performing Gaussian averaging of the three-dimensional material composition profile with spatial extent $\sigma = 0.9$~ nm. The schematic picture in Fig.~\ref{Fig1}(c) shows a surface of constant indium concentration with corners smoothed by this averaging.

We minimize the elastic energy of the system within the theory of continuous elasticity on a uniform Cartesian grid to find the strain field. In a non-centrosymmetric material, the shear strain at the material interfaces induces a piezoelectric field, which we calculate to the second order in the strain-tensor elements. We use a state-of-the-art implementation~\cite{GawareckiPRB2014} of the multiband ${\bm k}\cdot{\bm p}$ theory in the envelope function approximation~\cite{BahderPRB1990} to find the eigenstates of electrons and holes in a QD. This calculation includes the strain, the piezoelectric field, the spin-orbit interaction, and the external fields. The explicit form of the Hamiltonian can be found in Ref.~\cite{Mielnik-PyszczorskiPRB2018}, while details of the QD modeling and material parameters used are given in Ref.~\cite{GawelczykPRB2017} and references therein. By numerically diagonalizing the Hamiltonian, we obtain the single-particle energy levels and carrier eigenstates as discretized pseudospinors of envelope functions for each subband.

Next, we use the configuration-interaction approach to calculate the exciton (bound electron-hole pair) and negative trion (two electrons and a hole) states with a configuration basis constructed of 12 electron and 12 hole states. Taking into account the Coulomb interaction and phenomenological electron-hole exchange interaction (corresponding to 60~\textmu{}eV splitting of bright exciton states), we obtain the exciton and trion eigenstates. Next, within the dipole approximation~\cite{Andrzejewski_2010}, we calculate the interband optical transition dipole moments for the lowest-energy bright transitions.
For the exciton, the dipole moment is given by
\begin{equation}\label{eq:kp-dipole}
    {\bm d} = -\sum_{jl}\frac{i  \hbar e}{m_0E_X}c_{jl}\left\langle \mathbf{\Phi}_l \right\rvert {\bm P} \left\lvert \mathbf{\Phi}_j \right\rangle,
\end{equation}
where $m_0$ is the electron mass, $E_X$ is the exciton (optical transition) energy, $\mathbf{\Phi}_{j(l)}$ are eight-component pseudospinors of electron envelope functions for the $j$th conduction ($l$th valence) single-particle eigenstates spanned in the standard eight-band ${\bm k}\cdot{\bm p}$ basis, $c_{jl}$ are the coefficients for expansion of the exciton eigenstate in the electron-hole configuration basis, and ${\bm P} = (m_0/\hbar)\partial H_{{\bm k}\cdot{\bm p}}/\partial {\bm k}$ is the momentum operator defined with respect to the ${\bm k}\cdot{\bm p}$ Hamiltonian.
Out of the two bright exciton states in the ground-state manifold, we choose the higher-energy one for further analysis. This is based on its polarization properties having more regular dependence on QD geometry and magnetic field.
Similarly, for the transition between the trion state and a given $m$th electron eigenstate,
\begin{equation}\label{eq:kp-dipole-trion}
    {\bm d} = -\sum_{jl}\frac{i \hbar e }{m_0(E_T+E_1-E_m)}
        \left( c_{jml} - c_{mjl} \right)
        \left\langle \mathbf{\Phi}_l \right\rvert {\bm P} \left\lvert \mathbf{\Phi}_j \right\rangle,
\end{equation}
where $E_T$ is the trion state energy, $E_m$ is the $m$th electron eigenstate energy (so that $E_1$ is the ground-state single-particle energy), and $c_{jml}$ are the coefficients for expansion of the trion eigenstate in the configuration basis.

We consider a single, selected transition and hence assume the QD to be a two-level system. The in-plane part of the interband dipole moment characterizing the relevant transition in the QD is parametrized as
\begin{equation}\label{dipol2}
{\bm d} = d_0
{\begin{pmatrix}
\cos(\chi_{\rm d}) \cos(\Psi_{\rm d}) - i\sin(\chi_{\rm d}) \sin(\Psi_{\rm d}) \\
\cos(\chi_{\rm d}) \sin(\Psi_{\rm d}) + i\sin(\chi_{\rm d}) \cos(\Psi_{\rm d})
\end{pmatrix}},
\end{equation}
consistently with the parameterization of the field polarization defined in Eq.~\eqref{eq:field}, with $d_0$ denoting the magnitude of the dipole moment and with the same geometrical interpretation of the parameters (see Appendix~\ref{AppElipsy}).

\subsection{Coupling}

The emission rate of the quantum emitter into a given eigenmode within the dipole approximation is given by 
\begin{align}
\gamma_{\bm k} \propto |{\bm d}^*\cdot {\bm e}_{\bm k} |^2\,,
\end{align}
where ${\bm e}_{\bm k}$ is the electric field of the guided mode with wave vector $\bm{k}$. In the presence of a transverse field component, ${\bm e}_{{\bm k}} \neq {\bm e}_{- {\bm k}}$. Therefore, the emission rates into counter-propagating modes are in general not equal ($\gamma_{{\bm k}}\neq \gamma_{-{\bm k}}$), resulting in a nonzero chirality factor defined as the normalized difference between the forward and backward emission rates
\begin{equation}\label{derectionalty}
D = \dfrac{\gamma_{{\bm k}} - \gamma_{-{\bm k}}}{\gamma_{{\bm k}}+\gamma_{-{\bm k}}}.
\end{equation}
As the light-matter coupling depends on the polarization state of the emitter, one is able to engineer this coupling by tuning the emitter polarization parameters.

An important advantage of photonic crystals is that they can enhance the emission from a quantum emitter into a desired guided mode due to the broadband Purcell effect and strong suppression of the coupling to radiation modes by the photonic bandgap effect~\cite{bykov1975spontaneous, lodahl2004controlling, yablonovitch1987inhibited,wang2011mapping, koenderink2006spontaneous}. Only QDs that simultaneously exhibit chiral coupling and high Purcell enhancement may fully benefit from chiral quantum optics. The Purcell factor reads~\cite{hughes2004enhanced}
\begin{equation}\label{eq:purcell}
F_{{\bm k}} ({\bm r})=  \dfrac{3\pi c^2 a n_{\rm g} (k)}{ \omega_{k}^2 \sqrt{\varepsilon({\bm r})}} |\hat{\bm{d}}^{*} \cdot  \hat{\bm{e}}_{\bm k}({\bm r})|^2,
\end{equation}
where $\hat{\bm{d}}=\bm{d}/d_0$, $\hat{\bm{e}}_k=\bm{e}_k/\tilde{e}_{k0}$ with the normalization $\tilde{e}_{k0} = \big[\int_{\text{SC}} d^3{\bm r} \varepsilon({\bm r}) | {\bm e}_{\bm k}({\bm r}) |^2\big]^{1/2}$, $\omega_k$ is the mode frequency, $\varepsilon ({\bm r})$ is the relative electric permittivity at the point $\rm{r}$, SC indicates integration over a single supercell, and the group index for the considered mode is given by
\begin{equation}
    n_g(k) = \dfrac{2c (U_{e,{\bm k}} +U_{h,{\bm k}}) }{\left\lvert\int_{\text{SC}} d^3{\bm r} {\rm Re} [ {\bm e}^*_{\bm k}({\bm r}) \times {\bm h}^*_{\bm k}({\bm r})]\right\rvert},
\end{equation}
where $c$ is the speed of light in vacuum, ${\bm h}_{\bm k}$ is the magnetic field of the mode, and $U_{e,{\bm k}}(U_{h,{\bm k}})$ is the time-averaged electric (magnetic) field energy in the super-cell volume. As a consequence of the enhancement of emission to a desired mode and suppression of coupling to radiation modes, a near-unity efficiency of emission into the desired guided mode (the $\beta$ factor) is achievable~\cite{arcari2014near}, even in the fast light regime~\cite{scarpelli201999}, and it is remarkably robust with respect to the position of the emitter in the waveguide~\cite{arcari2014near}. 

To determine the average coupling of the propagating mode with a QD emitter with a given polarization at a random in-plane position, we average the directionality over the area of the waveguide core. Since QDs cannot lie within the air holes and efficient coupling to the guided mode is only possible in the waveguide area, we restrict the averaging to the vicinity of the core indicated by the blue shaded region in Fig.~\ref{Fig1}(b). We consider half of the core region because the same coupling in the other half can be obtained by inverting all polarizations (note that averaging over the full core region formally always results in a vanishing average directionality). Furthermore, to suppress couplings with transverse magnetic modes, we place the quantum emitter in the symmetry plane $z= 0$ (dashed line in Fig.~\ref{Fig1}(b)).

\section{Results}\label{sec:results}

\subsection{Waveguide modes}\label{subsec:pcw}

\begin{figure}[tb]
	\includegraphics[width=\columnwidth]{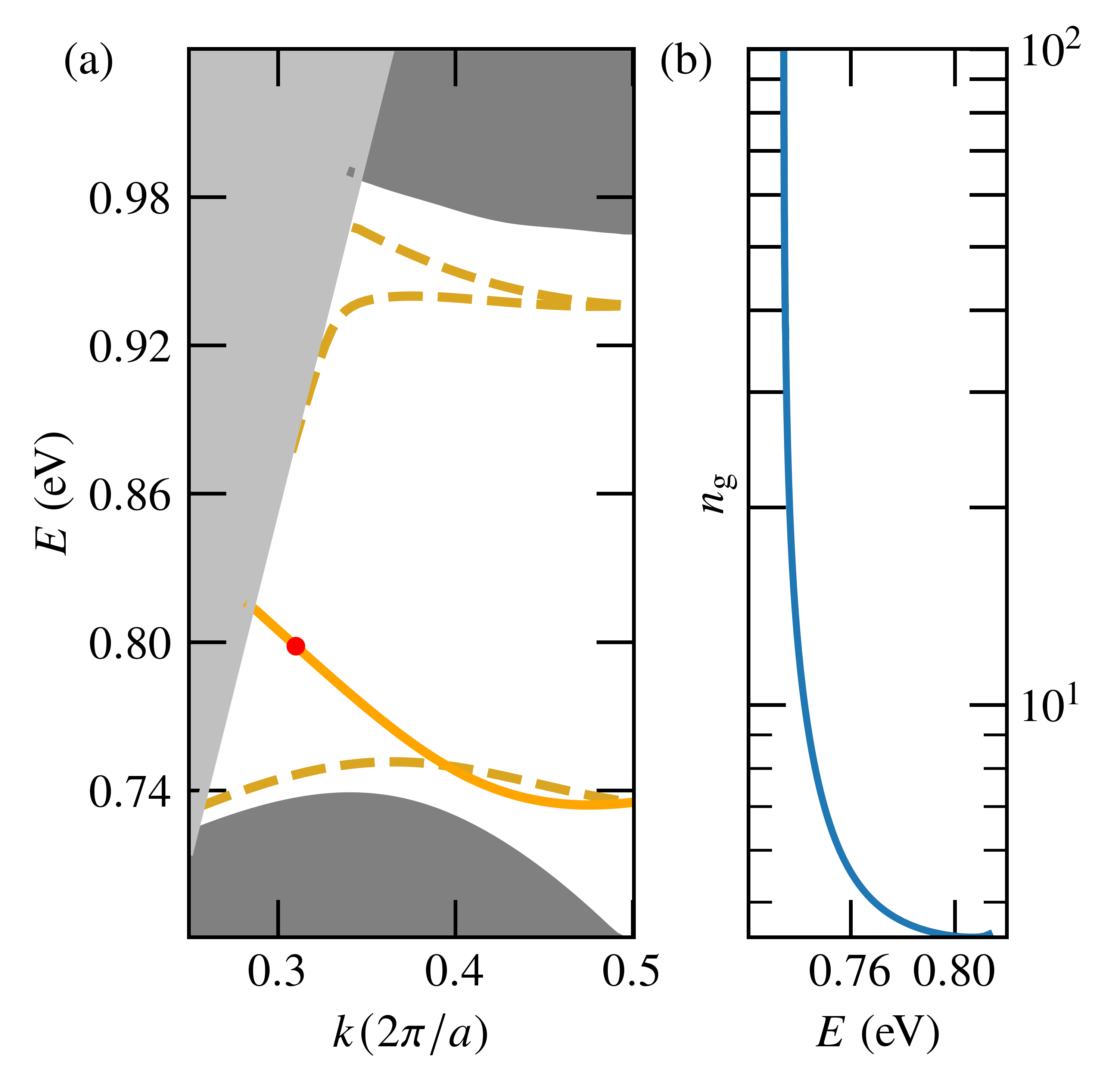}
	\caption{Projected TE band structure of the PCW displaying wave\-guide modes (orange). The gray regions mark the membrane guided modes (dark gray) and radiation modes (light gray) that are not bound to the membrane. The solid line indicates the guided mode considered in our study, whereas the dashed curves represent other, potentially leaky modes. The wavenumber chosen for further investigation is highlighted by the red dot. (b) Group index $n_{\rm g}$ of the considered guided mode as a function of mode energy. }\label{Fig2}
\end{figure}

Using the finite element method, we first determine the photonic TE-like bands, shown in Fig.~\ref{Fig2}(a), and the corresponding group indices [Fig.~\ref{Fig2}(b)] for the photonic crystal waveguide.
The defect modes are plotted with lines, whereas the extended modes in the photonic crystal and the extended modes propagating in the air (radiation modes) are represented as dark- and light-gray-shaded areas, respectively. We see that the chosen structure supports one well-confined mode plotted with a solid line. In addition, there are three modes near the bulk modes (dashed) that are anticipated to leak into the continuum in practice. The favorable guided mode spans a wide frequency range, offering a good spectral adjustment to the QD transition. In addition, it is a well-confined mode and, therefore, is selected for further consideration. 

 \begin{figure}[tb!]
	\includegraphics[width=\columnwidth]{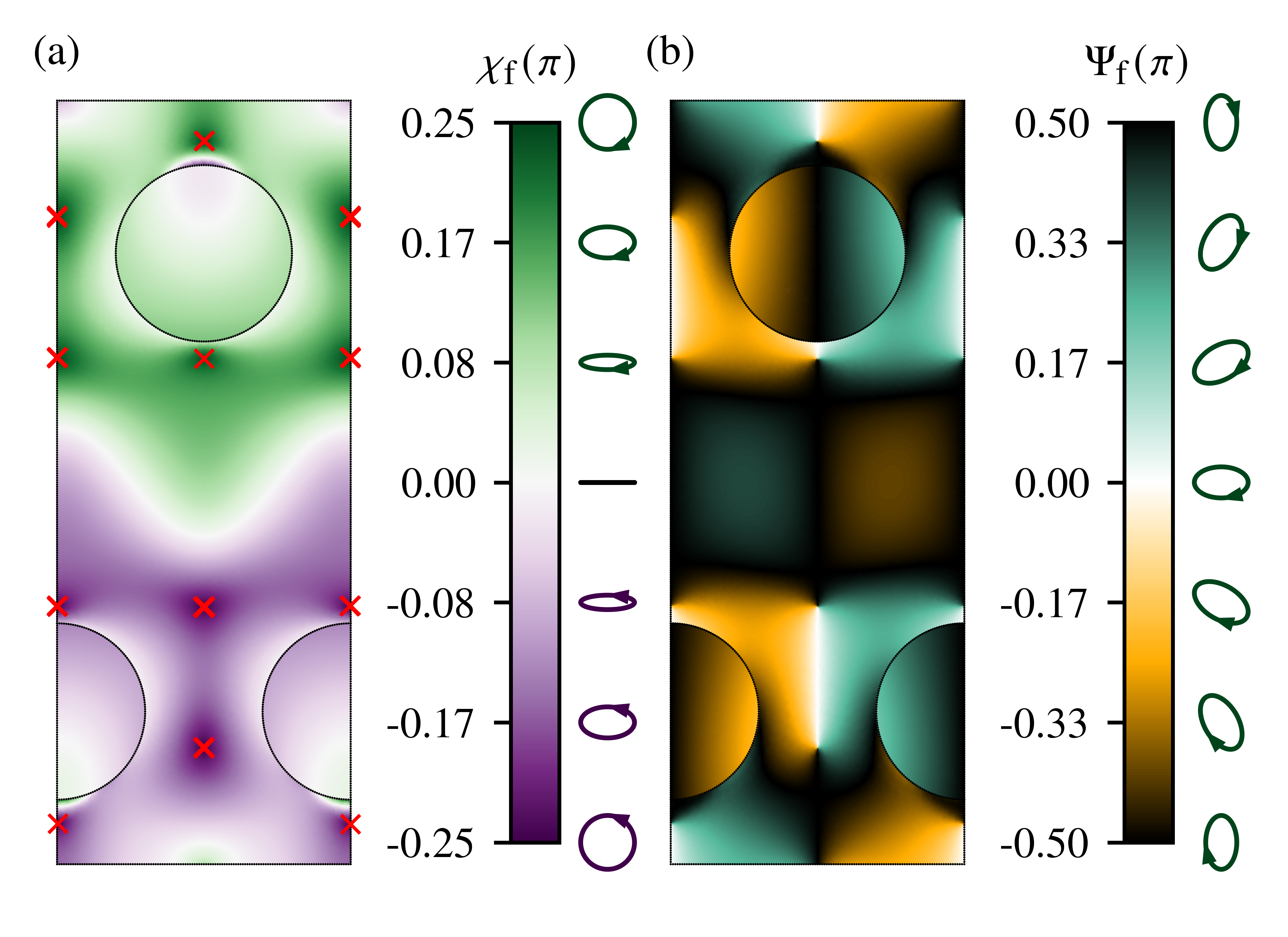}
	\caption{ Polarization ellipticity angle $\chi_{\mathrm{f}}$ (a) and orientation angle $\Psi_{\mathrm{f}}$ (b) within the core area of the PCW unit cell in the $z=0$ plane for the considered mode. States having right- and left-handed ellipses are shown in green and purple, respectively. The red crosses in (a) indicate C points with circular polarization.}\label{Fig3}
\end{figure}

Next, we calculate the polarization properties of the selected mode and extract the polarization parameters as described in Appendix~\ref{AppElipsy}. We show the ellipticity angle $\chi_{\rm f}$ in the $z = 0$ plane as a function of position in Fig.~\ref{Fig3}(a) and the orientation angle $\Psi_{\rm f}$ in Fig.~\ref{Fig3}(b). We can clearly see in Fig.~\ref{Fig3}(a) that each of the two helicities (right-handed in green and left-handed in purple) appears almost exclusively in one half of the waveguide. While the white line in the plot indicates linear polarization, the very dark spots, highlighted by the red crosses, mark circular polarization (C points). Apart from these tiny regions, part of which lie close to the air holes, the selected mode shows predominantly elliptical polarization. This again illustrates the potential of elliptical polarization for flexible device design ~\cite{lang2022perfect}.
Figure~\ref{Fig3}(b) shows the orientation of the polarization ellipse. 

\subsection{Emitter coupling}\label{sec:qds}

 \begin{figure}[tb!]
	\includegraphics[width=\columnwidth]{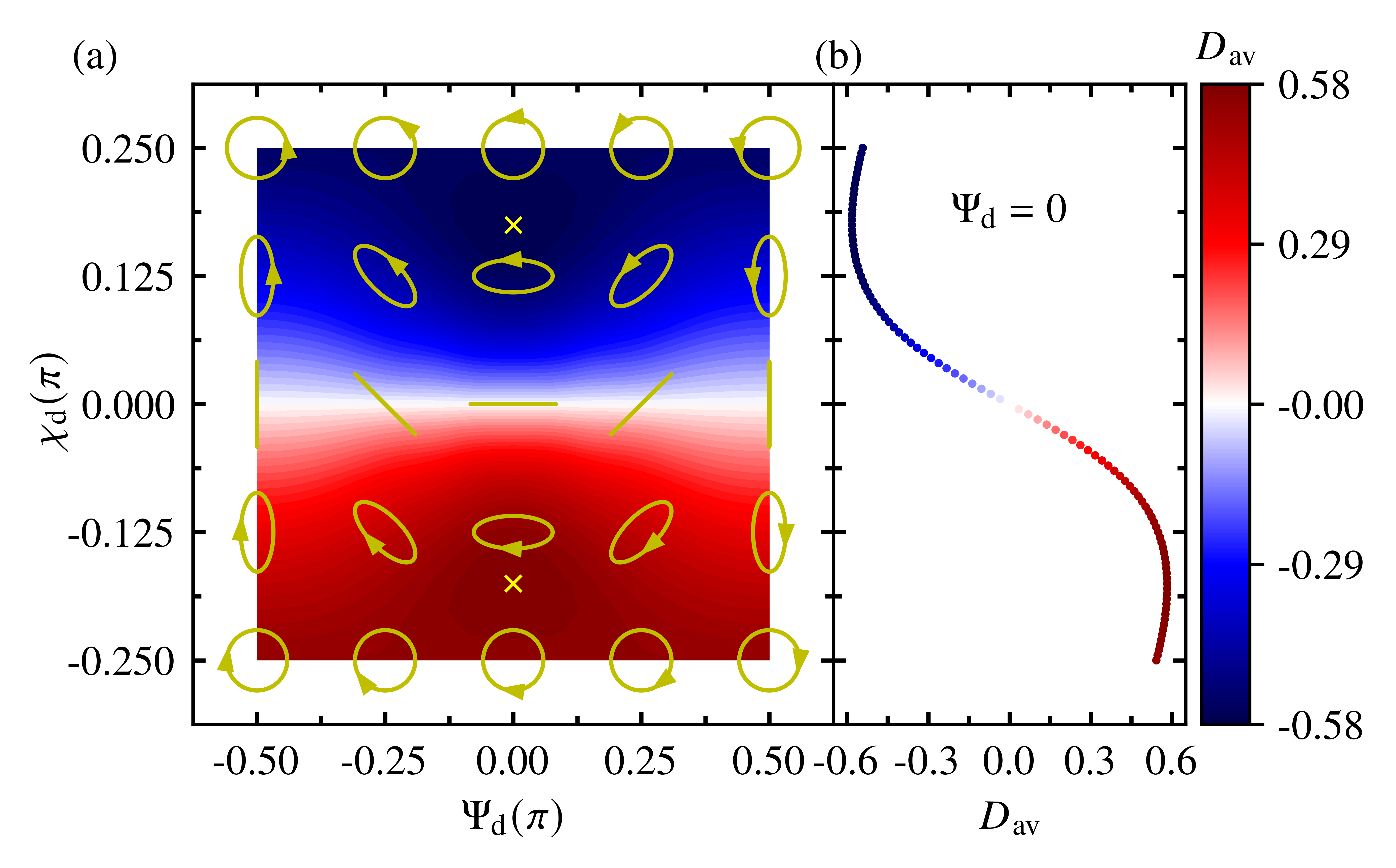}
	\caption{(a) Map of the averaged directionality of the emission from a randomly placed QD as a function of the polarization of the transition dipole moment parametrized by orientation angle $\Psi_{\rm d}$ and ellipticity angle $\chi_{\rm d}$. Yellow lines illustrate polarization states for a given $\Psi_{\rm d}$ and $\chi_{\rm d}$, while the yellow crosses indicate the maximum and minimum values of average directionality. (b) Cut line of the map in (a) for the orientation angle $\Psi_{\rm d} = 0$. }\label{Fig4}
\end{figure}

\begin{figure}[tb!]
	\includegraphics[width=\columnwidth]{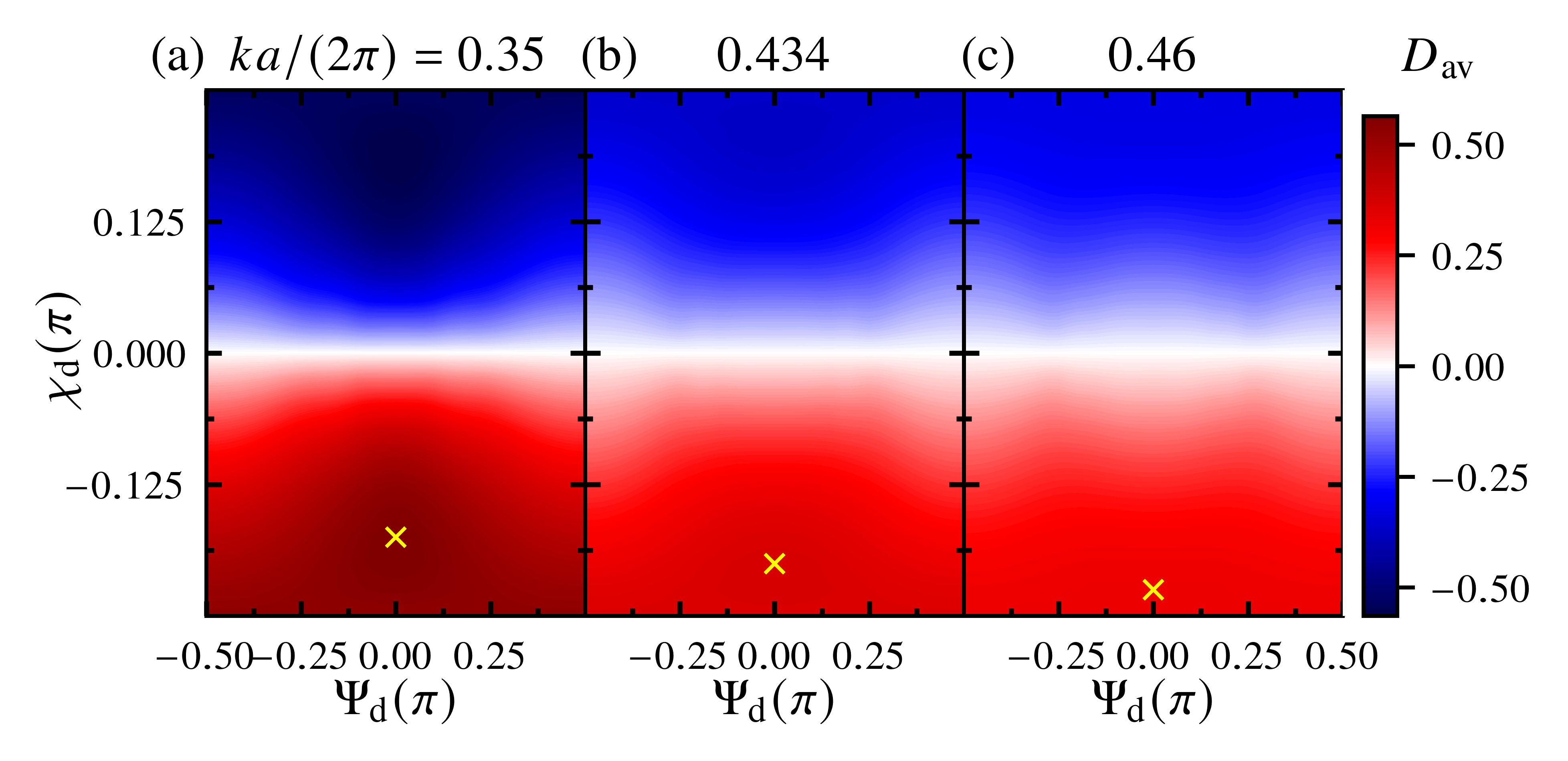}
	\caption{Average directionality map of QDs with random in-plane spatial positioning, as in Fig.~\ref{Fig4}(a) for (a) $k a/(2\pi) = 0.35$ ($n_{\text{g}} \approx 4.85$), (b) $k a/(2\pi) = 0.434$ ($n_{\text{g}} \approx 15.1$), (c) $k a/(2\pi) = 0.46$ ($n_{\text{g}} \approx 46.8$). Yellow crosses indicate the maximum values of average directionality.}\label{Fig5}
\end{figure}

We now consider a QD emitter coupled to the electric field of the selected guided mode. In the first step, we treat the emitter formally, characterizing its polarization by the orientation and ellipticity angles ($\Psi_{\rm d},\chi_{\rm d}$). For every polarization state of the dipole transition moment, we calculate the directionality from Eq.~\eqref{derectionalty} at a given point in the structure. As previously shown ~\cite{lang2022perfect}, with an appropriately polarized emitter, perfect directionality can be reached at a given position. However, in view of the random placement of the QD, a more informative figure of merit appears to be the directionality averaged over the QD position. Therefore, we average the result over half the core region [blue shaded region in Fig.~\ref{Fig1}(b)], as discussed above. Fig.~\ref{Fig4} presents the average directionality of the QD emission for different polarization states of the dipole transition moment of the QD for $ka/(2\pi) = 0.31$ ($n_{\text{g}} \approx 4.4$, red dot in Fig.~\ref{Fig2}(a)). The average directionality varies with both angles, reaching a maximum value of $\pm$0.58 for the optimal elliptical polarization with ($\Psi_{\rm d},\chi_{\rm d}) = (0,\mp0.178\pi$) and zero for all linear polarizations. The optimal parameter values correspond to the in-plane dipole $\hat{\bm{d}}=(0.848,\mp0.53i)$, which is the ellipse oriented along the propagation direction. 

Interestingly, the average directionality decreases with increasing group index. This is illustrated in Fig.~\ref{Fig5}, where we present the average directionality calculated for different wave numbers corresponding to increasing values of $n_{\mathrm{g}}$. As the light slows down from (a) to (c), the maximum average directionality decreases and the optimal ellipticity angle tends to approach circular polarization. Moreover, it is well known that in-plane backscattering between the counter-propagating modes scales as $\displaystyle n^2_{\rm g}$~\cite{hughes2005extrinsic}, leading to large losses in the slow light regime. Since for the chosen mode the group index increases with wave number $k$, both these effects favor low $k$. Therefore, we choose a possibly small wave number while staying away from the light cone to prevent coupling to the radiation modes. 

\begin{figure}[tb!]
	\includegraphics[width=\columnwidth]{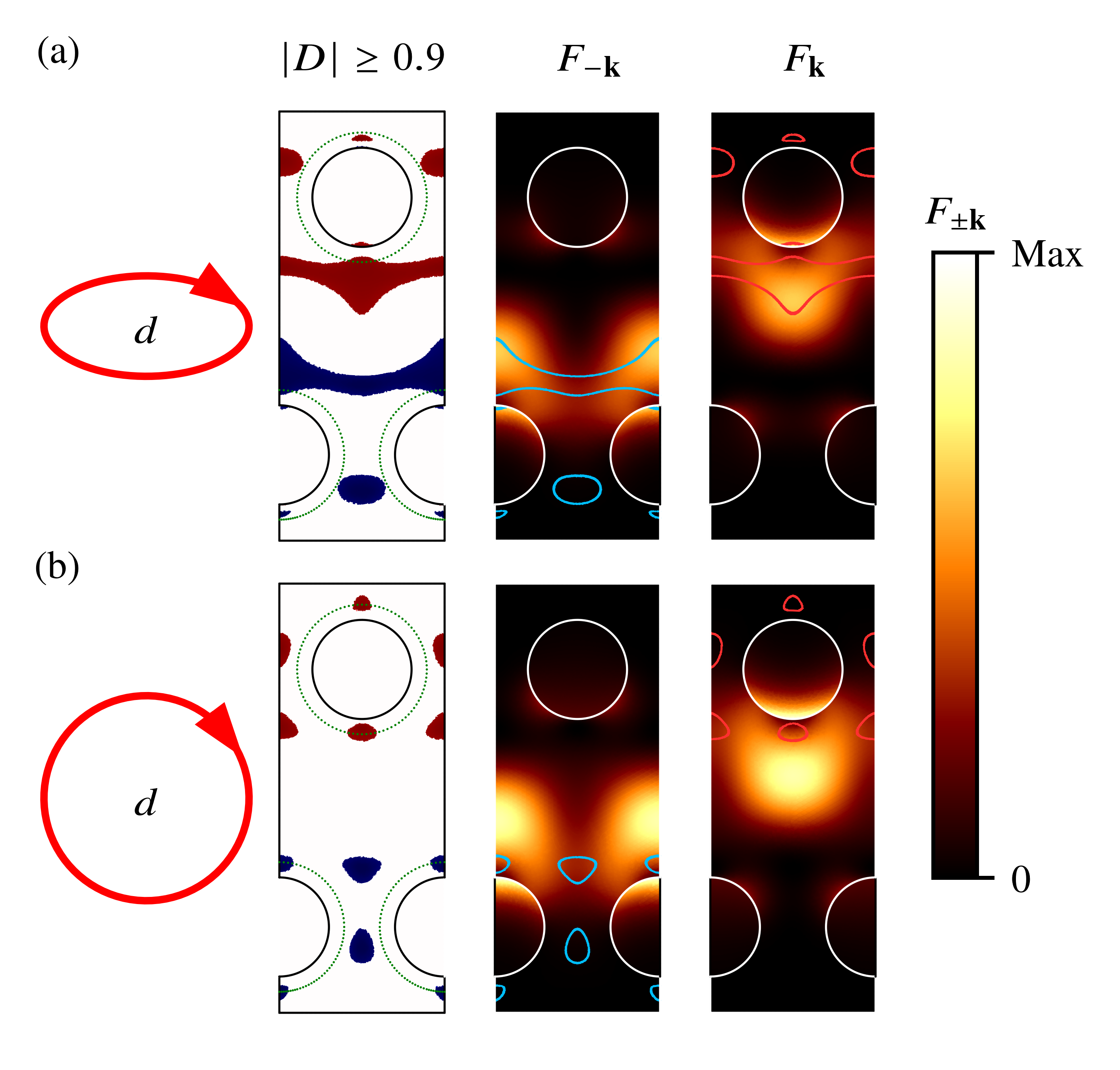}
	\caption{Areas where $|D| \geq 0.9$ (left) and Purcell factor (middle, right) for an optimized elliptical polarization state (a) and a circular one as a reference (b). In the left panels, the red areas indicate positive values of $D$, whereas blue ones denote negative values of $D$. The dotted green lines indicate a distance of 40~nm from the air-dielectric interfaces of the holes. In the middle and right panels the lines limiting the area where $|D| \geq 0.9$ are plotted onto the Purcell maps to illustrate the overlap between these two quantities.}\label{Fig6}
\end{figure}

We further calculate the directionality of emission $D$ from Eq.~\eqref{derectionalty} as a function of the emitter position, assuming the optimal transition dipole found above, as well as for the circular dipole for comparison, and mark the areas where $|D| \geq 0.9$. The calculated maps are shown in the left panels of Fig.~\ref{Fig6}(a) and~\ref{Fig6}(b) for the optimal dipole found above (with $\chi_{\rm d} = -0.178\pi$) and for a circular dipole (with $\chi_{\rm d} = -0.25\pi$), respectively. In the left panels, the areas with $|D|\geq 0.9$ are marked by red (propagation in $+x$ direction) and blue (propagation in $-x$ direction) colors, and we can directly see that in the case of elliptical polarization, two large regions spread continuously throughout the unit cell in the light propagation direction [Fig.~\ref{Fig6}(a)]. In contrast, for circular polarization in Fig.~\ref{Fig6}(b), the areas form smaller isolated spots. To quantify this difference, the overall area is about 4 times larger for optimized elliptical polarization than the one where a circular dipole is assumed, and more than 6 times larger than $A = \pi \xi^2$, where $\xi$ is the precision of embedding a QD in a photonic structure, which is $\xi\approx 40$~nm for InGaAs QD~\cite{pregnolato2020deterministic}. This shows that an emitter optimized with respect to the average directionality ensures a technological advantage in the manufacturing of the required structures.

Optimizing the dipole moment is beneficial only if high directionality can be combined with efficient emission. Therefore, in the middle and right panels of Fig.~\ref{Fig6}, we present maps of the Purcell factor calculated from Eq. \eqref{eq:purcell} for the optimal emitter. Although the magnitude of the Purcell factor is slightly reduced for the elliptically polarized dipole [middle and right panel in Fig.~\ref{Fig6}(a)] as compared to the circular dipole [middle and right panel in Fig.~\ref{Fig6}(b)], the area of considerable Purcell enhancement remains essentially unchanged. As a result, the optimized polarization yields a much more extended area where high directionality is combined with considerable Purcell enhancement. Another benefit of the optimal elliptical polarization that can be observed in the directionality map is a small shift of high directionality regions away from the air-dielectric interfaces of the holes, which is desired to sustain good optical quality of the QD emitter and compatibility with the current technology: Etching the holes affects their vicinity, which will alter the optical quality of the QDs in these areas. A 40~nm range around the air hole, consistent with current technology limitations~\cite{pregnolato2020deterministic}, is marked with a dotted green line in Fig.~\ref{Fig6}. This favorable shift observed for the optimal dipole further supports the relevance of the chosen figure of merit for the optimization.

\begin{figure}[tb!]
	\includegraphics[width=\columnwidth]{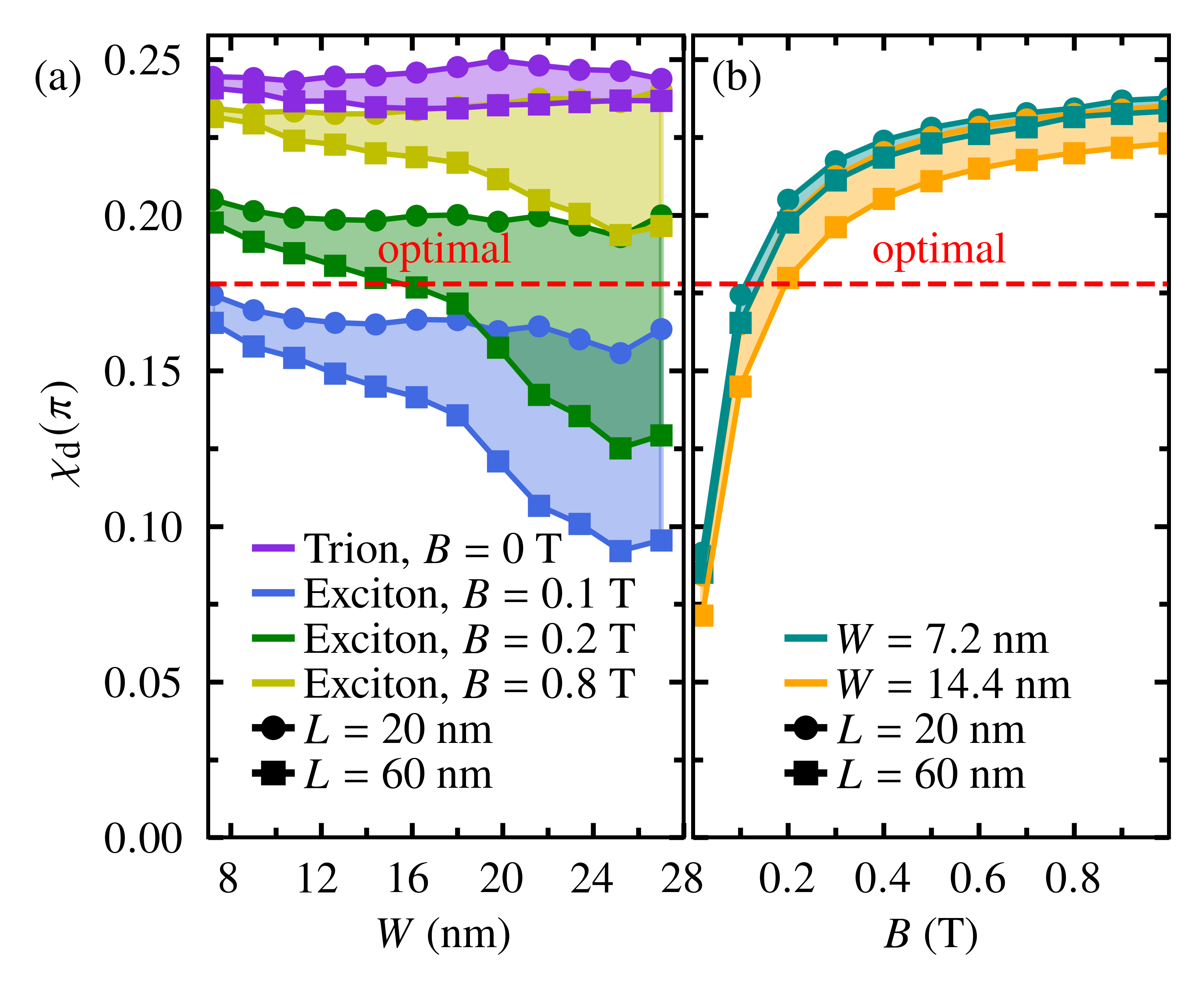}
	\caption{(a) Ellipticity angle of the polarization of the dipole transition moment in a QD as a function of QD width, illustrating both the neutral exciton and trion (violet).  In the case of an exciton, three distinct results are shown: for low magnetic fields of 0.1~T (blue) and 0.2~T (dark green), and a high magnetic field of 0.8~T (light green). (b)  Ellipticity angle of the polarization of the dipole transition moment in a QD as a function of applied magnetic field for a neutral exciton. Results are presented for QDs of varying widths, with dark cyan representing small widths and orange for larger widths. In both (a) and (b), data are presented for QDs of two different lengths: $L=20$~nm (circles) and 60~nm (squares). The red dashed line indicates the optimal ellipticity angle found in Fig.~\ref{Fig4}.}\label{Fig7}
\end{figure}

Finally, we show that the optimal value of the transition dipole is feasible with neutral exciton transitions within realistic QD characteristics. For a series of QDs described in Sec.~\ref{subsec:qd} with different geometrical parameters, we calculate the single-particle, exciton, and trion eigenstates, as well as their optical transition dipole moments. In Fig.~\ref{Fig7}, we present the analysis of the polarization properties of QDs, specifically focusing on the ellipticity angle of the transition dipole. In Fig.~\ref{Fig7}(a), we study the ellipticity angle as a function of QD width, for two values of the QD length and a fixed height-to-width ratio, showing results for both neutral excitons and trions. For excitons, our analysis reveals a noteworthy finding: we can achieve the optimal ellipticity angle in a weak magnetic field regime [blue color in Fig.~\ref{Fig7}(a)] for QDs with larger widths. However, these ellipticity angles are not achievable with trions [violet color in Fig.~\ref{Fig7}(a)]. In Fig.~\ref{Fig7}(b), we further investigate the dependence of the ellipticity of the transition dipole on the applied magnetic field for neutral excitons. The behavior is qualitatively the same for all geometries, with a minor impact of QD width, and with the length playing a larger role in the case of wider (higher) QDs. To achieve the situation in which the polarization of the QD transition matches the optimized ellipticity (marked as a dashed red line) we find that a magnetic field of $B\approx 0.15$~T is sufficient for realistic QD lengths of a few tens of nm.

\begin{figure}[tb]
	\includegraphics[width=.8\columnwidth]{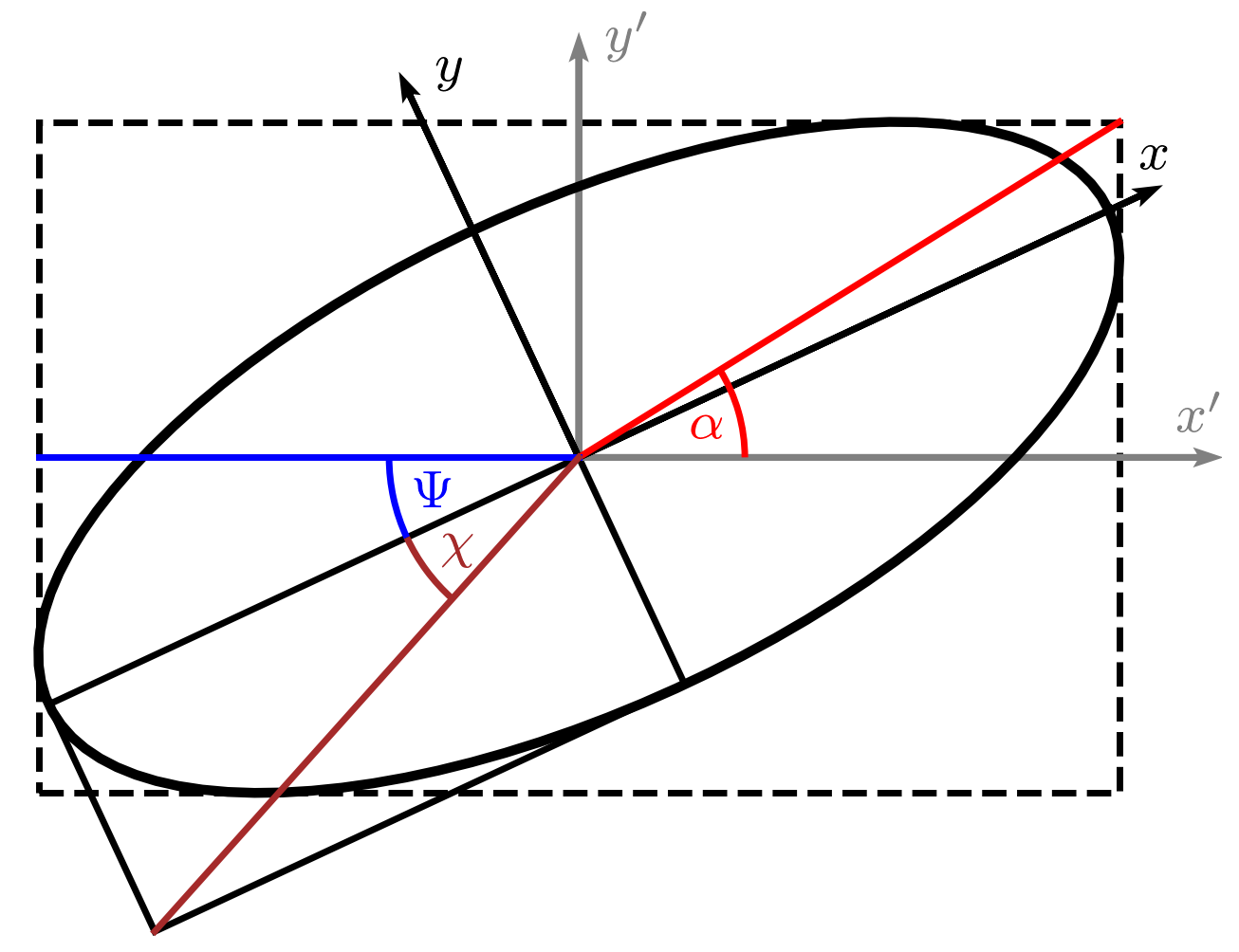}
	\caption{An illustration of the polarization ellipse with indicated parametrization. The parameter $\delta$ does not have a straightforward geometrical representation.}\label{Fig8}
\end{figure}

\section{Conclusions}\label{sec:conclusions}

We have studied the possibility of using a QD emitter for enhancing the chiral coupling to a guided light mode in a glide-plane waveguide. We choose the spatially averaged directionality as our figure of merit. We show that an emitter optimized in this way, apart from maximizing the overall chiral coupling, shows favorable properties from the point of view of the structure manufacturing: It enhances the area of the waveguide core where the directionality is high as compared to circularly polarized emitters and offers improved overlap between regions of high directionality and large Purcell enhancement \cite{lang2022perfect}. 
Using the ${\bm k}\cdot{\bm p}$ method, we have further demonstrated that optimal polarization properties can be achieved within realistic compositional and morphological characteristics of QDs by exploiting neutral exciton transitions at weak magnetic fields. 

Thus, the use of appropriately designed QDs for elliptically polarized emission is a potential path to improved scalability for quantum information processing and communication devices based on chiral interfaces, partly overcoming the challenges posed by the lack of precise control over lateral QD positions in epitaxial QD fabrication, which has hindered the achievement of efficient chiral interfaces for realistic QD ensembles so far. 
This finding presents an important step towards achieving robust chiral coupling in QDs-based photonic integrated circuits. 

\acknowledgements
M.\,G. is grateful to K. Gawarecki for sharing his ${\bm k}\cdot{\bm p}$ code.
The authors thank Ben Lang for useful suggestions regarding the initial version of the manuscript.
Part of the calculations has been carried out using resources provided by the Wroclaw Centre for Networking and Supercomputing (http://wcss.pl), Grant No. 203 and Grant No. 521. D.\,W. acknowledges the financial support of Science Foundation Ireland (SFI) under Grant numbers 18/RP/6236 and 22/PATH-S/10656.

\appendix

\section{Polarization Ellipse}\label{AppElipsy}

The polarization state of a mode at a given point or of an optical transition dipole moment is described in terms of the geometrical parameters of the polarization ellipse. Two different parameterizations given by angle pairs $\Psi,\chi$ and $\alpha,\delta$ can be used~\cite{goldstein2017polarized} to describe the polarization. The meaning of those angles is shown in Fig.~\ref{Fig8}. The ellipticity angle $\chi$ ($-\pi/4 \geq \chi \geq \pi/4$)  is defined as a ratio of the length of the minor semiaxis of the ellipse $u$ to the length of its major semi-axis $v$, such that $\tan(\chi) = u/v$. The orientation (also known as tilt or azimuth angle) $\Psi$ ($-\pi/2 \geq \Psi \geq \pi/2$) is the angle between the major axis of the ellipse and the $x$ axis, which defines the orientation of the ellipse in its plane. In the alternative parameterization, $\delta = \delta_y - \delta_x$ ($0 \geq \delta \geq 2\pi$) is the phase difference between two components of the electric field. and $\alpha$ ($0 \leq \alpha \leq \pi/2$) is defined in terms of the Cartesian components of the field amplitude as $\tan(\alpha) = e_{0y} /e_{0x}$. The polarization is right-handed (RH) if the ellipse is traversed in a clockwise sense when looking against the propagation direction (looking ``into the beam'', which corresponds to looking towards the page in Fig.~\ref{Fig8} if the electromagnetic wave propagates in the positive $z$-direction). The RH (LH) polarization corresponds to positive (negative) values of the ellipticity angle.

From the simulation, we obtain the electric field amplitudes of the modes $e_{0x}$ and $e_{0y}$, as well as the phase difference $\delta$ between them. Next, we calculate the Stokes parameters
\begin{equation}
    \begin{aligned}
        S_0 &=e_{0x}^{2}+e_{0y}^{2},\\
        S_1 &=e_{0x}^{2}-e_{0y}^{2},\\
        S_2 &=2e_{0x} e_{0y}\cos(\delta), \\
        S_3 &=2e_{0x} e_{0y}\sin(\delta). \\
    \end{aligned}
\end{equation}
Finally, we determine the ellipticity $\chi$ and orientation $\Psi$ angle from the Stokes parameters
\begin{equation}
    \begin{aligned}
        \chi&= 0.5 \arcsin \left({S_3}/{S_0} \right), \\
        \Psi &= 0.5\arctan \left( {S_2}/{S_1} \right).
    \end{aligned}
\end{equation}
Alternatively, the dipole can also be characterized using a second set of angles ($\alpha$, $\delta$),
\begin{equation}\label{dipol}
{\bm d} =
    \begin{pmatrix}
        \cos(\alpha) \\
        \sin(\alpha)e^{i\delta}
    \end{pmatrix}.
\end{equation}
In this case, we can extract the ellipticity and orientation angle from the relations
\begin{equation}
    \begin{aligned}
        \cos(2\alpha) &=\cos(2\chi) \cos(2\Psi), \\
        \tan(\delta) &= \tan(2\chi)/\sin(2\Psi).
    \end{aligned}
\end{equation}

\bibliography{paper}

\end{document}